%% file: main.tex
\documentclass[sigconf, noacm]{acmart}
\acmConference[arXiv Preprint]{}{May 2025}{Earth}

\AtBeginDocument{%
  }

\renewcommand\footnotetextcopyrightpermission[1]{ \footnotetext{ This is a
  preprint of an article submitted to a conference on planet Earth (mostly
  harmless), orbiting a small star in the uncharted backwaters of the
  unfashionable end of the western spiral arm of the Milky Way. \\
    © 2025 The Authors. Licensed under CC BY 4.0.
  }
}

\usepackage{tikz}
\usepackage{amsmath}

\usepackage{filecontents}

\input{common.tex}

\newif\ifprintcomments
\printcommentstrue  

\newcommand{\zhiyun}[1]{
  \ifprintcomments
    \textcolor{cyan}{{\em\bf Zhiyun:} #1}
  \fi
}

\newcommand{\work}{\mbox{\textsc{BugLens}}\xspace}

\begin{document}

\input{abbr.tex}

\title{The Hitchhiker's Guide to Program Analysis, Part II: 
Deep Thoughts by LLMs}
\author{Haonan Li}
\email{hli333@ucr.edu}
\affiliation{%
    \institution{UC Riverside}
    \city{Riverside}
    \state{California}
    \country{USA}
}

\author{Hang Zhang}
\email{hz64@iu.edu}

\affiliation{%
    \institution{Indiana University Bloomington}
    \city{Bloomington}
    \state{Indiana}
    \country{USA}
}

\author{Kexin Pei}
\email{kpei@cs.uchicago.edu}

\affiliation{%
    \institution{University of Chicago}
    \city{Chicago}
    \state{Illinois}
    \country{USA}
}

\author{Zhiyun Qian}
\email{zhiyunq@cs.ucr.edu}

\affiliation{%
    \institution{UC Riverside}
    \city{Riverside}
    \state{California}
    \country{USA}
}

\begin{abstract}
 Static analysis plays a crucial role in software vulnerability detection, yet
faces a persistent precision-scalability trade-off. In large codebases like the
Linux kernel, traditional static analysis tools often generate excessive false
positives due to simplified vulnerability modeling and over-approximation of
path and data constraints. While Large Language Models (LLMs) demonstrate
promising code understanding capabilities, their direct application to program
analysis remains unreliable due to inherent reasoning limitations.

We introduce \work, a post-refinement framework that significantly enhances
static analysis precision for bug detection. \work guides 
LLMs through structured reasoning steps to assess security
impact and validate constraints from the source code. When evaluated on
Linux kernel's taint-style bugs detected by static analysis tools, \work
improves precision approximately 7-fold (from 0.10 to 0.72), substantially
reducing false positives while uncovering four previously unreported
vulnerabilities. Our results demonstrate that a well-structured, fully-automated
LLM-based workflow can effectively complement and enhance traditional static
analysis techniques.
\end{abstract}

\settopmatter{printfolios=true}
\maketitle
\input{sec/intro.tex}
\input{sec/bg.tex}

\input{sec/design_nominted.tex}
\input{sec/impl.tex}
\input{sec/eval_nominted.tex}
\input{sec/end_nominted.tex}




\bibliographystyle{plain}
\bibliography{main}

\clearpage
\input{sec/appendix_nominted.tex}

\end{document}

%% file: common.tex
\usepackage{amsmath,amsfonts}
\usepackage{algorithm}
\usepackage{algorithmicx}
\usepackage{algpseudocode}
\usepackage{graphicx}
\usepackage{textcomp}
\usepackage{listings}
\usepackage{xcolor}
\usepackage[english]{babel}
\usepackage{color,colortbl}
\usepackage{xspace}
\usepackage{multirow}
\usepackage{graphicx}
\usepackage{booktabs}
\usepackage{url}
\usepackage{xurl}
\usepackage{tipa}
\usepackage[T1]{fontenc}
\usepackage[utf8]{inputenc}
\usepackage{threeparttable}
\usepackage{amsmath,amsfonts}
\usepackage{pifont}
\usepackage{tabularx}
\usepackage[printonlyused]{acronym}
\usepackage[most]{tcolorbox}
\usepackage{multicol}
\usepackage{wrapfig}
\usepackage{adjustbox}
\usepackage{fontawesome5}

\usepackage{hyperref}
\hypersetup{
	pdftitle={},
	pdfauthor={},
	pdfsubject={},
	pdfkeywords={},
	bookmarksnumbered=true,
	bookmarksopen=false,
	colorlinks=true,
	pdfstartview=FitB,
	pdfpagemode=UseOutlines
}

\usepackage{tikz}
\usetikzlibrary{shapes,arrows,positioning,fit}
\usetikzlibrary{backgrounds}
\usetikzlibrary{arrows.meta, positioning, calc}

\definecolor{mygreen}{rgb}{0,0.6,0}
\definecolor{mygray}{rgb}{0.5,0.5,0.5}
\definecolor{mymauve}{rgb}{0.58,0,0.82}
\definecolor{pastelcyan}{rgb}{0.741, 0.867, 0.894}
\definecolor{skyice}{rgb}{0.62, 0.776, 0.953} 
\definecolor{warmlinen}{rgb}{0.949, 0.937, 0.906}
\definecolor{powderblue}{RGB}{152,193,217}
\definecolor{sageleaf}{RGB}{233, 237, 201}
\definecolor{blushcloud}{RGB}{241,231,231}
\definecolor{gainsboro}{RGB}{220,220,220}

\definecolor{mygold}{RGB}{175, 149, 0}   
\definecolor{mysilver}{RGB}{180, 180, 180} 
\definecolor{mybronze}{RGB}{173, 138, 86} 

\usepackage[font=footnotesize,skip=0pt,belowskip=0pt,aboveskip=0pt]{caption}
\usepackage{caption}
\newcommand{\etc}{\textit{etc.}\xspace}

\newcommand{\ie}{\textit{i.e.,}\xspace}    
\newcommand{\eg}{\textit{e.g.,}\xspace}    

\newcommand{\squishlist}{
	\begin{list}{$\bullet$} {
			\setlength{\itemsep}{0pt}
			\setlength{\parsep}{0pt}
			\setlength{\topsep}{0pt}
			\setlength{\partopsep}{0pt}
			\setlength{\leftmargin}{1.0em}
			\setlength{\labelwidth}{1em}
			\setlength{\labelsep}{0.5em}
		}
		}
		\newcommand{\squishend}{
	\end{list}
}

\newcommand{\find}[1]{
\begin{tcolorbox}[leftrule=1mm,rightrule=1mm,toprule=0mm,bottomrule=0mm,left=1pt,right=1pt,top=0.5pt,bottom=0.5pt]
\em #1
\end{tcolorbox}
}

\newcommand{\cut}[1] {}

\cut{\zhiyun{a few things we should do before submission:
		\begin{itemize}
			\item check captilization consistency, ``figure -> Figure'', caption of a figure/table should have the first letter captilized only
			\item center the caption of all figures/tables.
			\item remove the ``period'' mark in captions.
			\item spell check / grammar check (put all sentences in grammarly)
			\item for all the functions referened in textit\{\}, we should add the brackets. For example, ``textit\{fun\} -> textit\{fun()\}''
			\item global replace of ``\ textit'' with ''\ texttt''. Also, there is no need to add the quotation marks around \texttt.
			\item check the text that references the figure and the figure itself are always co-located on the same page (or one page before/after).
		\end{itemize}
	}}

%% file: abbr.tex
\newacro{LLM}{\textit{Large Language Model}}
\newacro{LLM4SA}{\textit{LLM for Software Analysis}}
\newacro{SecIA}{Security Impact Assessor}
\newacro{ConA}{Constraint Assessor}
\newacro{SAG}{Structured Analysis Guidance}
\newacro{PKA}{Project Knowledge Agent}
\newacro{AC-Hypo}{Aribitary Control Hypothesis}

%% file: sec/intro.tex
\section{Introduction}

Static analysis has long served as a cornerstone technique for identifying software
vulnerabilities.
By analyzing the code
without dynamic execution, these techniques aim to detect various security weaknesses, 
such as buffer overflows and information leaks. 
However, static analysis tools often struggle to balance the trade-off between
precision and scalability \cite{gosain_static_2015,DBLP:journals/csur/ParkLR22}.



More precise analysis, for example, symbolic execution \cite{horvath_scaling_2024}, 
can be computationally expensive and often infeasible for large codebases 
such as the Linux kernel~\cite{ubitect}. Conversely, more scalable techniques 
sacrifice the precision for scalability, leading to a high number of false positives.
For example, Suture~\cite{zhang_statically_2021}, an 
advanced taint bug detection in the Android kernel
shows a 90\% raw false positive rate, which requires manual
inspection of the results.

Specifically, its imprecision stems from two main issues:
\squishlist

\item \textbf{Simplified Vulnerability Modeling.} Static analyzers
may often rely on \textit{heuristic} simplified rules for vulnerability detection. 
For example, a static analyzer may
flag every \textit{arithmetic operation} as \textit{potentially overflowing}.
While this simplication might ensure no genuine vulnerabilities are missed, it
inflates the number of false positives.

\item \textbf{Over-Approximation of Path and Data Constraints.} In
order to avoid exponential path exploration, static analyzers often make coarse
assumptions about whether a path is feasible or how data flows through the
program. This over-approximation ensures analysis completes in a reasonable
time, but it also flags numerous \textit{infeasible} paths as potentially vulnerable,
resulting in excessive false positives.

\squishend

\noindent
Recent advances in \textit{Large Language Models }(LLMs) offer a promising
 avenue for overcoming these issues. Trained on vast amounts of code and natural
 language, LLMs exhibit remarkable capabilities in understanding code semantics,
 API usage patterns, and common vulnerability types \cite{khare_understanding_2024,
 stoica_if_2024, fang_large_2024}. 
 By leveraging these
 broader insights, LLM-based approach might able to: (1) \textit{enhance
 vulnerability modeling} by providing a more nuanced understanding of code
 semantics, and (2) \textit{refine path and data constraints} by a selective
 analysis of semantically plausible 
 paths and data flows.


However, LLMs are \textbf{not a silver bullet} for program analysis. Despite their
semantic understanding capabilities, LLMs are not inherently equipped for the
rigorous demands of program analysis \cite{qian_lamd_2025,he_code_2024}. Their
reasoning proves brittle, particularly when confronted with the complex program
dependencies crucial for security analysis \cite{kambhampati_can_2024,
chen2025reasoning, chapman_interleaving_2025}. Indeed, our initial experiments
confirm that naively applying LLMs to program analysis, for instance, by simply
asking \textit{``Is this warning a true positive?''}, yields highly unreliable results,
frequently misclassifying vulnerabilities or failing to identify critical flaws.
This is often because LLMs tend to fixate on surface-level code features,
missing the critical dependencies that dictate program behavior and
security properties, especially within intricate control and data flows.

Our work addresses this challenge by introducing a structured guidance framework
that directs LLM reasoning according to static program analysis paradigm. This
approach rests on the premise that static analysis, while theoretically sound,
is limited by practical tradeoffs in precision and scalability. By situating
LLM reasoning within this established methodology, we \textit{``compel''} 
the model to analyze more rigorously than it would by default, thus mitigating the inherent
limitations of LLMs in code reasoning.



In this work, we introduce \work, an innovative framework that
\textbf{\textit{post-refines}} the results of static analysis using LLMs. Rather
than blindly applying LLMs to analyze program, \work is carefully orchestrated
to teach LLMs key concepts of program analysis and guide them toward reasonable
analytical procedures. By analyzing the output of static analysis, \work\
complements the limitations of existing tools, especially in terms of precision,
and yields more accurate and actionable vulnerability detection for practical
codebases. We demonstrate that this combined approach significantly improves the
precision of taint-bug detection in the Linux kernel, reducing the need for
manual inspection of false positives, and even uncovering previously 
ignored vulnerabilities.

We summarize our contributions as follows:

\squishlist
    \item \textbf{Post-Refinement Framework.} We introduce \work, a
    post-refinement framework that supplements traditional static analysis to
    boost precision of the results, overcoming various practical weaknesses
    identified from real-world static analysis tools.

    \item \textbf{\acf{SAG}.} We design a structured workflow that 
        directs LLMs to follow the principles and processes of 
        established static analysis paradigms. This guided approach achieves 
        better vulnerability detection compared to naive LLM prompting strategies.
    
    \item \textbf{Empirical Results.} Evaluated on the Linux kernel,
    our solution improved the precision of a real-world static analysis tool
    from 0.1 to 0.72. Interestingly, prior manual analysis incorrectly filtered
    four warnings that were retained by \work.

    \item \textbf{Open Source.} We open-source our code and data of \work 
    on GitHub\footnote{\url{https://github.com/seclab-ucr/BugLens}}.

    

    
\squishend

%% file: sec/bg.tex
\section{Background \& Motivation}
\label{sec:moti}


\subsection{Taint Bugs}
Taint-style bugs involve insecure data propagation, where unsanitized data (taint \textit{source}) reaches a sensitive program location (taint \textit{sink}), causing flaws like out-of-bound access. For instance, an unchecked input integer (source) might become an array index (sink). Proper \textit{sanitization} (\eg range checks) is typically missing or insufficient. Such bugs can lead to severe vulnerabilities like buffer overflows or denial-of-service.

In the kernel, these bugs often occur when untrusted user input (\eg syscall arguments) propagates to sensitive kernel operations (\eg arithmetic calculations) without adequate sanitization~\cite{machiry_dr_2017,zhang_statically_2021}. Detecting these bugs is challenging due to:

\squishlist
    \item \textbf{Diverse and General Sinks.} Unlike user-space applications with often well-defined sink APIs (\eg \texttt{exec()}), the kernel's high privilege means almost \emph{any} operation (even a single arithmetic calculation) could be a sink if affected by unsanitized user data. This broad scope increases false positives. Additionally, low-level coding optimizations common in the kernel (\eg intentional integer overflows) can complicate accurate detection.

    \item \textbf{Scattered and Intricate Sanitization.} Sanitization checks can be distant from sinks, often crossing function boundaries within the large and complex kernel codebase. These checks are frequently intertwined with domain-specific kernel logic (\eg privilege systems, configurations), posing significant hurdles for taint analysis tools.
\squishend

\subsection{Static Analysis for Taint Bugs}
Suture~\cite{zhang_statically_2021} is the state-of-the-art work from academia aiming at kernel taint-style bug detection, which extends the previous work, Dr. Checker~\cite{machiry_dr_2017}, by adding cross-entry taint tracking capability and improving analysis precision.
CodeQL~\cite{codeql2025} is one of the most powerful and widely used industry code analysis engines, capable of taint analysis~\cite{taint_tracking_CodeQL} and query-based bug detection based on customizable rules.
CodeQL has a large community that develops and maintains many different bug detection rules, including those for kernel taint-style bugs (\eg \cite{backhouse_stack_2018}).
Despite their effectiveness, all these tools exhibit high false alarm rate (\eg \textasciitilde 90\%), underscoring the inherent difficulties in accurately identifying kernel taint-style vulnerabilities.
We elaborate on these difficulties and challenges in \S\ref{sec:chal_and_rationale}.

\section{Challenges and Design Rationale}
\label{sec:chal_and_rationale}

\subsection{Challenges in Static Analysis}

\subsubsection{Challenge 1 (C1): Simplified Vulnerability Modeling}
\label{subsubsec:c1_revised}

The first source of imprecision is the reliance on \textit{simplified
vulnerability detection} modeling. 
To maintain scalability and avoid missing potential bugs, static analyzers often employ overly simplified detection rules.

For instance, a static analyzer typically flags specific code patterns as
potential bugs. 
A \textit{Tainted Arithmetic Detector (TAD)} would identify any arithmetic operation involving tainted variables (\eg \texttt{\colorbox{warmlinen}{var} += size}, where \colorbox{warmlinen}{\texttt{var}} is tainted) as a potential integer overflow vulnerability. 
However, in Linux kernel practice, these patterns frequently do
not translate to exploitable vulnerabilities for two key reasons:

\squishlist
\item \textbf{The Behavior Itself is \textit{Benign}}. The kernel's reliance on \textit{low-level C idioms, pointer manipulation, and intentional `unsafe' design patterns} often confounds traditional analyzers, leading to inaccuracies, e.g., our evaluation shows that 61.2\% of the flagged `bugs' were actually benign (\S\ref{subsubsec:secIA-effectiveness}). 
For example, 
in Linux drivers, there are cases where integer overflow is expected and benign, such as with \texttt{jiffies}:

\includegraphics[width=\linewidth]{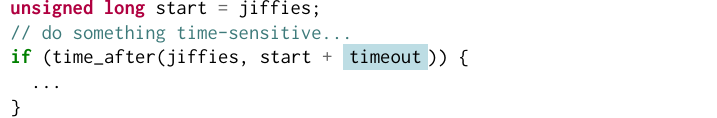}

For these cases, even when integer overflow is possible due to the tainted
variable \texttt{timeout}, it is not a security vulnerability, as \texttt{time\_after} is defined as \texttt{\#define time\_after(a, b) (b - a) > 0}).
Such
false positives commonly arise from intentional design patterns in the Linux
kernel, including struct hacks (deliberate out-of-bounds access), type casting,
unions, and data structure operations (\eg \texttt{ptr->next},
\texttt{container\_of}).

\item \textbf{The Tainted Data is Properly Validated} before reaching the sink.
For example, the \textit{Taint in the Loop Bound} (TLB) pattern is common in the
Linux kernel, but many include proper validation. Consider this array traversal:
\includegraphics[width=\linewidth]{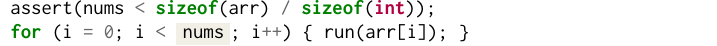}

Without the \texttt{assert} check, this loop could lead to an out-of-bounds
access. However, the \texttt{assert} ensures that the loop index never exceeds
the array bounds, making it safe. Static analyzers that
does not account for such checks (\eg path-insensitive ones) will misclassify this as a
vulnerability, leading to false positives.

\squishend


\subsubsection{Challenge 2: Over-Approximation of Constraints}
\label{subsubsec:c2_revised}

As mentioned above, the tainted data is often checked before the sink, and with a safe range.
However, static analysis tools often  \textit{over-approximate} the path and data constraints to ensure scalability, as the kernel's \textbf{immense scale and continuous evolution} make precise static analysis impractical, and the logic of the sanitization could often be distant or complex.
Alternatively, a common approach in static analysis \cite{infer_2015} is to directly prune paths with checks, regardless of their actual effects, which leads to false negatives. 
The existence of checks is not always effective in preventing the bug from happening, \eg we discovered some unreported bugs due to a plausible but ineffective check (see \S\ref{sec:eval_rq1}).
Consider the function \texttt{clamp\_num()} shown in the following, which is designed to sanitize the tainted variable \colorbox{pastelcyan}{\texttt{num}}:

\includegraphics[width=\linewidth]{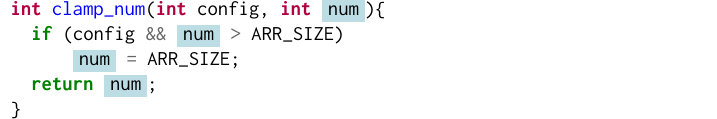}

The function \texttt{clamp\_num} only sanitizes the tainted variable 
\colorbox{pastelcyan}{\texttt{num}} effectively when the \texttt{config} is set.
Therefore, a precise static analysis needs to perform (i) a range analysis on the tainted variable \colorbox{pastelcyan}{\texttt{num}}, and (ii) a path-sensitive analysis on \texttt{clamp\_num()} to determine path conditions.

\subsection{The Opportunity in LLMs}





Recent advances in Large Language Models (LLMs) offer a promising avenue to
address these challenges, particularly in \textit{post-refining} the results
generated by scalable static analyzers. Trained on vast amounts of code and
natural language, LLMs exhibit remarkable capabilities relevant to overcoming
the limitations described above:

\squishlist
\item \textbf{Enhancing Vulnerability Modeling (Addressing C1):} LLMs possess a
nuanced understanding of code semantics, common programming patterns, and API
usage \cite{li2025codei, ni2024next, liu_generating_2025, khare_understanding_2024}. 
They can potentially
recognize complex or subtle vulnerability indicators that are difficult to
capture with predefined simplified heuristics. By analyzing code context (including
comments and variable names), LLMs might differentiate between genuinely
unsafe patterns and benign ones.

\item \textbf{Refining Path and Data Constraints (Addressing C2):} While not
performing formal reasoning, LLMs can leverage their semantic
understanding to assess the \textit{feasibility} of paths and data flows flagged by
static analysis. They might leverage contextual information 
to determine if the flagged bug is likely infeasible in practice,
thereby helping to prune false positives arising from coarse path and data constraint
analysis.
\squishend

\noindent
Recent works\cite{wen_automatically_2024,li_iris_2025} also show that LLMs can be
used to refine static analysis results (C1 \& C2) in a straightforward manner,
\ie to directly ask LLM with \textit{``Is this case from static analysis report vulnerable?''} followed by the code snippet. 

\subsubsection{Challenge 3: Reasoning Hurdles for LLMs.}
\label{subsubsec:c3}





Despite their potential, LLMs are not inherently suited for rigorous program
analysis tasks out-of-the-box. Their reasoning can
be fragile when dealing with complex program dependencies, and they lack
systematic constraint-solving mechanisms. 
Our experiments (detailed in \S\ref{sec:eval_rq2}) demonstrate that
such a simple prompting strategy leads to high false negatives --- cases where
the LLM incorrectly classifies actual vulnerabilities as safe code, especially
against the presence of complex control flow and data constraints.

We observe that the false negatives stem from the fundamental limitation of existing LLMs when tasked with sophisticated, structured, and symbolic reasoning. 
Specifically, with simple prompting, LLMs often rely on surface-level spurious features \cite{mccoy_embers_2024} to reason about the vulnerability. 
For example, the presence of a \textit{sanity check} often correlates statistically with \textit{safe code} (many practical static analyzers use such heuristics to prune paths). 
We observe the model classifies code containing such a check as safe without performing the deeper reasoning required to determine \textit{if the check is actually effective under the current execution} paths. 
Real-world vulnerabilities often exploit exactly these scenarios: checks that are bypassable, incomplete, or rendered ineffective by intricate control and data flows. 
The model's tendency to rely on spurious patterns without performing the actual reasoning has been shown ineffective in tracking critical dependencies and results in unreliable predictions in many different domains \cite{prabhakar_deciphering_2024}.

\subsection{Design Rationale}

To address challenges C1 through C3, we introduce \work, a fully automated, multi-stage LLM-based framework designed specifically as a post-refinement layer for static analyzers. 
The key design philosophy of \work is to scale the test-time compute of LLMs by allocating more tokens for precise reasoning of the intended program behaviors based on the model's understanding of program semantics~\cite{snell2024scaling}.
Specifically, \work consists of the following specific key components, each targeting a specific challenge:

\squishlist
  \item \textbf{\acf{SecIA}}: \textit{Addressing Simplified
  Vulnerability Models (C1)}. 
  Instead of relying solely on predefined patterns, \work leverages LLMs to analyze the \textit{potential security impact} if tainted data identified by static analysis were completely controlled by an attacker.
  It then evaluates whether the tainted value could potentially lead to 
  security vulnerabilities (\eg memory corruption, denial of service (DoS)), as detailed in \S\ref{subsec:bug-eval}. 
  By focusing on semantic consequences rather than simplified vulnerability modeling, \ac{SecIA} provides a more precise assessment of security impact. 

  \item \textbf{\acf{ConA}}: \textit{Addressing Over-Approximation
of Path \& Data Constraints (C2).} 
ConA uses LLMs to analyze whether data constraints in the code are sufficient to prevent potential vulnerabilities from being triggered. By tracing how tainted data is processed and constrained, ConA performs heuristic reasoning (\eg to find sanitizations with names \texttt{check\_x})  to evaluate the effectiveness of these safeguards. This approach is designed to handle the complexity and scalability challenges of systems like the Linux kernel, where traditional formal methods may lack precision or scalability.

\item \textbf{\acf{SAG}}: \textit{Addressing Reasoning
Hurdles for LLMs (C3).} 
As described in \S\ref{subsubsec:c3}, simply prompting LLMs for analysis often leads to spurious reasoning.
To this end, SAG grounds LLMs' reasoning in \ac{ConA} with the scaffold that describes the typical program analysis workflow. 
At a high level, \ac{SAG} employs specific prompts to describe key analysis principles and include in-context examples to demonstrate how to systematically dissect code, trace dependencies, and evaluate conditions, especially in complex scenarios. 
This guidance constrains the LLM towards a more rigorous and reliable analysis process.
\S\ref{subsec:llm-guide} elaborates on the design of \ac{SAG}.

\squishend

%% file: sec/design_nominted.tex
\begin{figure*}
  \centering
  \hspace{-10pt}
  \includegraphics[width=.95\textwidth]{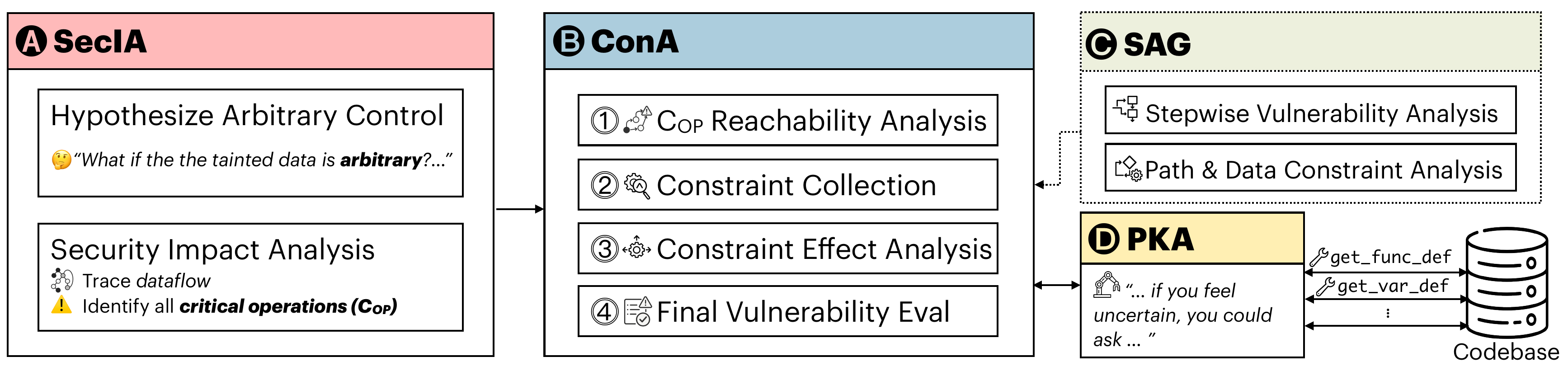}
  \caption{Overview of \work, 
  showing (A) \acf{SecIA} first assesses the security impact 
  of the potential bugs identified by static analysis, and then (B) \acf{ConA}
  assesses the data constraints and evaluates if the bug is feasible. 
  ConA is guided by (C) \acf{SAG} to reason on the code more effectively,
   and  can interact with the (D) \acf{PKA} to
  get information about the codebase on-demand.}
  
  \label{fig:wf-1}
  \Description{Overview of our workflow, showing (1) static analysis and (2) LLM-based filtering}
\end{figure*}

\section{Design}
\label{sec:design}

\work first takes the static analysis report as input, which identifies
the context of the potential bugs and the tainted data flow,
and then performs the following steps to evaluate the potential bugs,
as shown in Figure~\ref{fig:wf-1}:

\squishlist
  \item \textbf{\acf{SecIA}}: This component evaluates the security impact of the potential bugs
    identified by static analysis. It identifies the Critical Operations (C\textsubscript{Op}) that
    are influenced by the tainted data and classifies them as either
    \textit{Normal Code} or \textit{Requires Constraint Analysis}. 
  \item \textbf{\acf{ConA}}: This component performs a multi-step analysis to evaluate the feasibility of the potential bugs. It collects the path conditions and data constraints, summarizes them, and evaluates whether they are effective in preventing the vulnerability.
  \item \textbf{\acf{SAG}}: This component provides the scaffold to constrain the LLM through pre-defined code reasoning procedures to help it understand the code and analyze the constraints effectively. 
  \item \textbf{\acf{PKA}}: This component allows the LLM to access the codebase on-demand, enabling it
    to retrieve global codebase information.
\squishend

Additionally, \work also adopts the common prompting techniques to scale the test-time compute, such as
(1) {Majority-vote querying}, where we query the model
multiple times and take the most common answer; 
(2) {Chain-of-Thought (CoT)
prompting}~\cite{wei_chain--thought_2023}; 
and 
(3) {Schema-constrained summarization}, where a follow-up prompt requests the model's own output in a predefined XML format, making LLM's response easy to parse.

\subsection{ \acf{SecIA} }
\label{subsec:bug-eval}

The core insight behind \textit{\ac{SecIA}} is based on the fundamental evaluative question:
\textit{What are the consequences if tainted data assumes arbitrary values?}

\subsubsection{Core Assumption and Rationale}
\ac{SecIA} operates on a fundamental assumption regarding attacker capability at
the point of initial impact assessment: \textit{\textbf{\acf{AC-Hypo}}}. For a
given program location \(K\) where static analysis reported an operation
\(Op(v)\) involving tainted data \(v\) (the \textit{sink}), (1) we hypothesize that
an attacker can control \(v\) to take any value \(v_{atk}\) (within the
constraints of its data type), and (2) we \textit{provisionally ignore} any
 effects of checks or path conditions (even explicit checks)
encountered on analysis. In other words, we assume that the attacker 
can take \textit{any} value to anywhere (successors of the sink node in the 
control flow graph).

This hypothesis enables SecIA to streamline analysis. 
SecIA focuses solely on
\textit{potential security impact}. 
This permits \textit{early filtering} of findings based purely on the consequence, (safely) reducing subsequent analysis load.
Critically, this approach \textit{defers} the complex analysis of actual
program constraints—including path feasibility, value ranges, and importantly,
\textit{whether those protective checks or sanitizers}
really work.
This deferral \textit{mitigates
false negatives} (FNs) by preventing premature dismissal of vulnerabilities due
to reliance on potentially bypassable checks or inaccurate LLM constraint
reasoning about their effectiveness. The effectiveness is shown in
\S\ref{subsubsec:secIA-effectiveness}.

\subsubsection{Security Impact Analysis}

SecIA performs a \textit{forward influence analysis} to identify the
influenced critical operations (\textit{C\textsubscript{Op}}) that are
potentially affected by the tainted data \(v\) from the sink location \(K\).
It then filters out the benign operations that are not security-sensitive based on whether this operation could (potentially) result in memory bugs (\eg out-of-bound access, arbitrary memory access) or DoS. 
For instance, the \texttt{jiffies} case in \S\ref{subsubsec:c1_revised} will be attributed as ``not a bug''.

\begin{figure*}
  \centering
  \hspace{-10pt}\includegraphics[width=.98\textwidth]{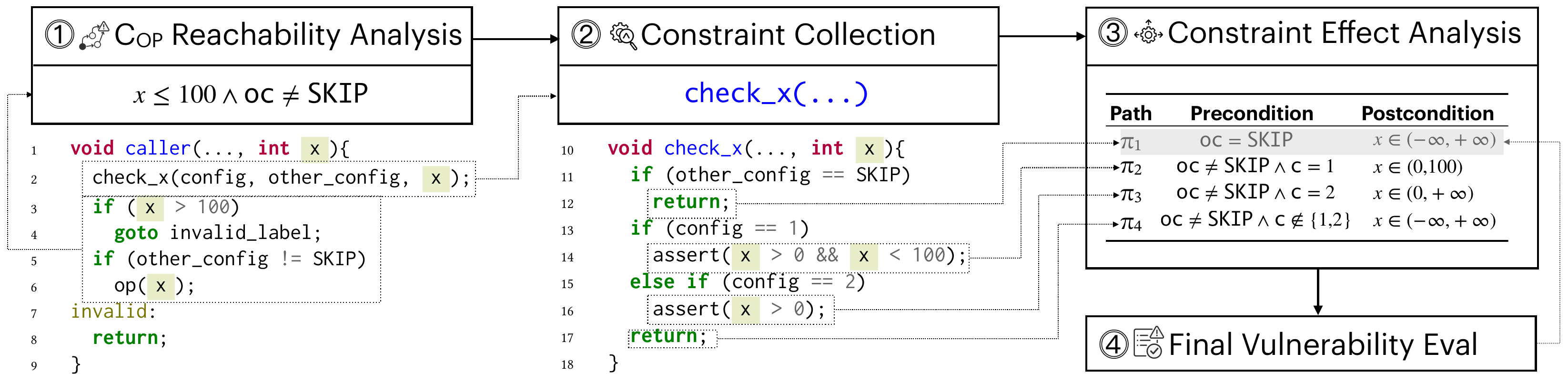}
  \caption{ 
    The workflow of \acf{ConA} with an example, where the \texttt{x} is tainted
  and \texttt{op(x)}
   is the critical operation (C\textsubscript{OP}). The sanitization function \texttt{check\_x(..., x)}
  is affected by the \texttt{config} (noted as \texttt{c}) and \texttt{other\_config} (noted as \texttt{oc}).}
  \Description{This figure illustrates the multi-stage workflow of  \ac{ConA}.}
  \label{fig:ci_wf_case}
\end{figure*}

\subsection{\acf{ConA}}
\label{subsec:san-finder}

The \textit{\acf{ConA}} aims to identify whether the bug of C\textsubscript{Op} can be triggered by analyzing the data constraints. 
As shown in Figure~\ref{fig:ci_wf_case}, the Constraint Assessor involves a four-step workflow:

\squishlist
    \item \textbf{Step 1: Critical Operation Reachability Analysis.} The process begins by determining the conditions under which program execution can reach the specific \textit{Critical Operation (C\textsubscript{Op})} location. 
     This establishes the base requirements for the vulnerability to be possible.

    \item \textbf{Step 2: Constraint Collection.} 
    Next, the analysis traces the tainted data flow path(s) backward from the C\textsubscript{Op} towards the data's source. Along this path, it identifies code segments—such as conditional statements, assertions, or calls to validation functions—that appear intended to act as \textit{constraints} on the tainted data's value or range before it is used at the C\textsubscript{Op}.

    \item \textbf{Step 3: Constraint Effect Analysis.}
    Each potential constraint identified in Step 2 is then analyzed in detail. This step aims to understand the constraint's specific \textit{effect}: Under what conditions (\textit{precondition}) does the constraint need to satisfy, and what impact does it have on the tainted variable's possible numerical range (\textit{postcondition})?

    \item \textbf{Step 4: Final Vulnerability Evaluation.} Finally, 
    it performs an evaluation to determine if the potential vulnerability 
    can still be triggered. If yes, the case is classified as a
    \textit{{``Potential Vulnerability.''}}
    Otherwise, if the LLM determines that the constraints effectively prevent the vulnerability from being triggered, the case is classified as \textit{{``Eliminated.''}}
    
\squishend

\noindent
This workflow leverages LLMs to interpret code (mimicking a formal reasoning process), identify relevant patterns and constraints, and perform the reasoning required for each step. 

We employ this LLM-based approach acknowledging the trade-offs of soundness and precision.
\ac{ConA} aims to overcome the potential precision limitations of conservative abstractions and achieve broader applicability in the complex Linux kernel.
In \S\ref{sec:eval_rq1} and \S\ref{sec:eval_rq3}, we thoroughly examine how our design in relaxing the strict over-approximation introduces only a reasonably small number of false negatives.
In the following sections (\S\ref{subsubsec:san-step1} through \S\ref{subsubsec:san-step4}), we elaborate on the specific design, rationale, heuristic considerations, and soundness implications inherent in each stage of this analysis process.

\subsubsection{Step 1: Critical Operation Reachability Analysis}
\label{subsubsec:san-step1}
The analysis begins by determining the conditions required for program execution 
to reach the specific \textit{Critical Operation} (C\textsubscript{Op}) location previously 
identified by  \ac{SecIA}  as potentially vulnerability.
These reachability conditions form the base constraints that must 
be satisfied for the vulnerability to be triggerable 
via this C\textsubscript{Op}.

\noindent
\textbf{Condition Representation:} When analyzing path conditions, we allow the
LLM to generate natural language summaries that capture the semantic meaning of
complex reachability conditions. For instance, where traditional analysis might struggle,
an LLM might identify a condition like: 
\textit{``The operation is only executed if \texttt{init\_subsystem()} returned 
zero (success),
AND the device state in \texttt{dev->status} equals \texttt{STATUS\_READY}.''} 

The LLM is later asked  (in Step 4) to evaluate these semantic constraints to assess the potential impact. 
The natural language representation allows the LLM to leverage its understanding of the code's intent and context but introduces a higher degree of uncertainty in the final evaluation, which we must acknowledge.


\subsubsection{Step 2: Backward Constraint Collection}
\label{subsubsec:san-step2}
Once the Critical Operation (C\textsubscript{Op}) and its local reachability conditions are identified, 
the next step is to gather potential constraints imposed on the tainted data before 
it reaches the C\textsubscript{Op}. This involves tracing the data flow path(s) for the tainted 
variable backward from the C\textsubscript{Op} toward its source(s).

\noindent
\textbf{Example:} In Figure \ref{fig:ci_wf_case}, tracing back from \texttt{critical\_op(x)} in \texttt{caller()}, 
the LLM identifies 
the call \texttt{check\_x(...)} to be a potential constraint on \texttt{x}. 
It would query the PKA for the function definition and
 analyze its effect, as detailed in Step 3.

\subsubsection{Step 3: Constraint Effect Analysis}
\label{subsubsec:san-step3}

After collecting potential constraining code segments (like the function \texttt{check\_x()}) in Step 2, this step analyzes the effect of these segments on the tainted variable. 
The goal is to understand how different execution paths within these segments modify the possible range of the tainted variable and under which conditions (\textit{preconditions}) those paths are taken.

This step prompts the LLM to first identify all major execution paths through the provided code segment (\eg \texttt{check\_x()}), and then considers the preconditions and postconditions (data constraints of the tainted data) of each path.

\noindent
\textbf{Example:} In Figure \ref{fig:ci_wf_case}, 
we ask the LLM to analyze the function \texttt{check\_x()}
 to summarize its effects on the tainted variable \texttt{x}.
Particularly, the path \(\pi_1\) and \(\pi_4\) are \textit{\textbf{bypass}} paths, 
and \texttt{check\_x()} will not effectively limit the 
range of \texttt{x} with these preconditions.

\subsubsection{Step 4: Final Vulnerability Evaluation}
\label{subsubsec:san-step4}
This final step synthesizes the precondition and postcondition from the previous analyses to determine if the identified constraints 
effectively neutralize the potential vulnerability associated with the Critical Operation (C\textsubscript{Op}). And then it classifies the reported vulnerability as either ``\textit{Eliminated}'' (constraints effectively prevent the vulnerability happening) or ``\textit{Potential Vulnerability}'' (no effective constraints).

\noindent
\textbf{Example:} 
In Figure~\ref{fig:ci_wf_case}, Step~1 of the reachability analysis shows that \texttt{oc}~$\neq$~\texttt{SKIP}, and therefore the path $\pi_1$ can be eliminated, as its precondition is \texttt{oc}~$=$~\texttt{SKIP}. Since \texttt{c} and \texttt{oc} are unknown, the function \texttt{check\_x()} does not impose additional range constraints (we can only assume path $\pi_4$ to be valid with conservative analysis).
The range analysis for \texttt{x} yields $(100, \texttt{INT\_MAX})$, and the final evaluation requires an understanding of \texttt{op()} itself. 
If the range is sufficient to prevent the bug, the issue is considered \emph{eliminated}.

\subsection{\acf{SAG}}
\label{subsec:llm-guide}

To support the fine-grained code reasoning steps in \ac{ConA}, we employ
\acf{SAG} to scale the test-time compute of LLMs with structured reasoning
templates and few-shot examples. At a high level, SAG grounds the reasoning
procedures of LLMs with typical program analysis steps by eliciting more
reasoning tokens during inference. Specifically, SAG assists Constraint Assessor
(ConA) with the following two types of analysis: (i) \textit{Guided Stepwise
Vulnerability Analysis} with step-by-step instructions that decompose the
analysis of precondition and postcondition (as described in
\S\ref{subsec:san-finder}), and (ii) \textit{Guided Path Condition and Data
Constraint Analysis} to demonstrate how to analyze challenging data constraints
and path conditions from code.

\subsubsection{Guided Path Condition Analysis} 
\label{subsubsec:llm-guide-condition}

In the analysis of the path condition in Step 1 and Step 3 of \ac{ConA}
(\S\ref{subsubsec:san-step1} and \S\ref{subsubsec:san-step3}),
we guide the LLM using prompts designed to extract path conditions 
from the source code surrounding the \textit{operation} of interest.
For example, consider the path leading to the 
\texttt{op(x)} call within the \texttt{caller} function:
\begin{center}
\includegraphics[width=\linewidth]{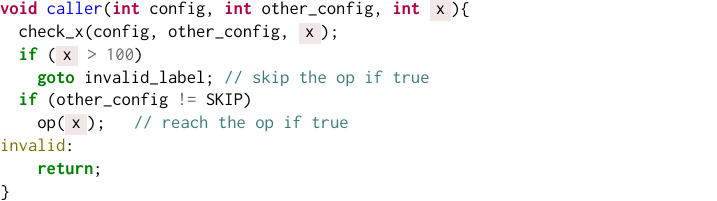}
\end{center}

\noindent
\ac{SAG} asks the LLM to identify and categorize:
\squishlist
  \item \textbf{Bypass Conditions:} Identify conditional statements 
  where taking a specific branch avoids the operation. The LLM is instructed to
   extract the condition and negate it to find the requirement for not 
   bypassing the operation.

  \textit{\textbf{Example}}: The condition \texttt{\colorbox{blushcloud}{x} > 100} leads to \texttt{invalid}, 
  bypassing \texttt{critical\_op(\colorbox{blushcloud}{x})}. 
  The negated condition required to proceed towards the OP is \texttt{x \(\le\) 100}.
  
  \item \textbf{Direct Conditions:} Identify conditional statements where taking a specific branch is 
  necessary to reach the operation along the current path. The LLM extracts the condition directly.
  
  \textit{\textbf{Example}}: Reaching sink(x) requires entering the if block, so the condition is \texttt{other\_config \(\neq\) SKIP}.
\squishend

\noindent
The LLM then combines these elementary conditions using logical AND to form the path-specific 
reachability constraint set for the operation. For this path in the example, the derived 
reachability condition is:
 \(x \leq 100 \land \texttt{other\_config != SKIP} \).


\subsubsection{Guided Data Constraint Analysis}
\label{subsubsec:llm-guide-constraint}

In the analysis of data constraints in Step 2 and Step 3 of \ac{ConA}
(\S\ref{subsubsec:san-step2} and \S\ref{subsubsec:san-step3}),
We consider the following data constraints:

\squishlist
  \item \textbf{Type constraints.}  
        The variable's static type already restricts its range
        (\eg \texttt{uint8} is always in the range of $[0, 255]$).
        
  \item \textbf{Validation (\textit{transferable} to source).}  
        The program \emph{tests} the value and aborts or
        reports an error if the test fails, \emph{without modifying the
        value}.  
        Because the check refers to the \emph{current value}, the
        knowledge gained from this check (\eg ``the value must be~$\ge 0$ on the
        success branch'') also applies to \emph{all source variables that influenced
        this value in the data flow}.  
        
  \item \textbf{Sanitization (\textit{not transferable} to source).}  
        The program \emph{writes a new, corrected value} back to the
        variable (\eg clamping it to a range).  
        This operation severs the connection to the original value, so
        any property we learn afterwards applies only to the sanitized
        copy, not to the original source variables in the data flow.
\squishend

The key difference is that \emph{validation knowledge travels
backward along the data-flow graph}, while sanitization overwrites the
flow and stops the transfer.

\begin{center}
\includegraphics[width=\linewidth]{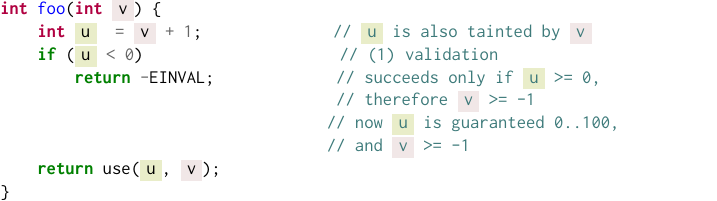}
\end{center}

Step (1) is a \emph{validation}: it \emph{reads} \texttt{\colorbox{sageleaf}{u}} and
branches, so the fact `\texttt{\colorbox{sageleaf}{u}}~$\ge 0$' (hence
\texttt{\colorbox{blushcloud}{v}}~$\ge -1$) becomes part of the path condition and is
\emph{transferable} to other variables in earlier nodes (\texttt{\colorbox{sageleaf}{u}=\colorbox{blushcloud}{v}+1}).  
Step (2) is a \emph{sanitization}: it \emph{writes} a new value into
\texttt{\colorbox{sageleaf}{u}}; the constraint “$0\le\texttt{\colorbox{sageleaf}{u}}\le 100$” holds only
\emph{after} this assignment. 
It worth noting that the sanitization to \texttt{\colorbox{sageleaf}{u}} does not 
pose any constraints to \texttt{\colorbox{blushcloud}{v}}. However, if we replace 
the \texttt{clamp()} with an \texttt{assert(\colorbox{sageleaf}{u}< 100)}, 
we would get a constraint of \texttt{\colorbox{blushcloud}{v}} as well, \( \texttt{\colorbox{blushcloud}{v}} < 99 \).

Formal program analysis naturally distinguishes between validation and sanitization; however, LLMs often confuse these concepts (\eg missing  transferred validation, treating sanitization as validation).
SAG emphasizes these differences with few-shot examples, and therefore 
improves the LLMs' reasoning on data constraints.

\subsection{\acf{PKA}}
\label{subsec:llm-agent}


\work leverages \acf{PKA}, a simple LLM code agent equipped with custom tools to automatically navigate the codebase and retrieve the related code context during the analysis.
It starts the analysis with the context of the sink and explicitly requests more context (\eg function definitions \texttt{get\_func\_def}, global variable definition \texttt{get\_var\_def}, struct layouts) if needed. 
PKA then retrieves the relevant code snippets iteratively until the model self-decides that it has sufficient context to complete the analysis.

We provide a set of request types, enabling the LLM to gather broader codebase information at any point. 
This design is both flexible and extensible: new request types can be
supported by adding corresponding backend callbacks.
PKA searches the relevant code snippets based on CodeQuery
\cite{kopathy_ruben2020codequery_2025}.
We implement PKA in 500 LOC of Python code.

%% file: sec/impl.tex
\section{Implementation}
\label{sec:impl}

\work is implemented with approximately 7k tokens (30k characters) in prompts
(detailed in \S\ref{apdx:prompt}) and 2k lines of Python code that manages API
requests and codebase querying functionality.





\subsection{Inferring Variable Names}


Static analysis tools typically operate on compiler-generated IR code (like LLVM IR), 
which differs significantly from source code representation. Since \work analyzes source 
code directly, we need to map variables between these representations.

The IR code provides line numbers that correspond to locations in the source code, 
along with data flow information. However, compiler optimizations often create, 
eliminate, or transform variables, making this mapping non-trivial. For example, 
LLVM might convert a simple ternary expression into a complex bitwise operation.

We leverage LLMs to perform this variable name inference by providing them with:
(1) the line number from source code and 
(2) the data flow from IR code. 
The LLM then identifies the corresponding source-level variables, 
effectively bridging the gap between IR-level analysis and source code understanding.

%% file: sec/eval_nominted.tex
\section{Evaluation}
\label{sec:eval}

\subsection{Research Questions}

Our evaluation aims to address the following research questions.

\squishlist
    \item \textbf{RQ1: (Effectiveness)} How effective is \work in identifying vulnerabilities?
    \item \textbf{RQ2: (Component Contribution) } How does the the prompt design affect the performance of \work?
    especially for the \ac{SecIA} and \ac{SAG} component?
    \item \textbf{RQ3: (Model Versatility)} How does the performance of \work vary across different LLMs?
\squishend

\subsection{Experimental Setup}
\label{sec:eval_setup}


We primarily evaluate \textit{\work}
using OpenAI's o3-mini model, specifically \texttt{o3-mini-2025-01-31}. 
This model was chosen as our primary focus because, as demonstrated in our evaluation for RQ3 (Section~\ref{sec:eval_rq3}), 
it achieved the best overall performance on our bug analysis task compared to several other recent leading models.

To address RQ3 regarding the generalizability of \textit{\work}, we also tested
its performance with a range of prominent alternative models, encompassing both
closed-source and open-source options. These include:
OpenAI's o1 (\texttt{o1-2024-12-17}), GPT-4.1 (\texttt{gpt-4.1-2025-04-14}),
 Google's Gemini 2.5 Pro 
(\texttt{gemini-2.5-pro-preview-03-25}), Anthropic's Claude 3.7 Sonnet 
(\texttt{claude-3-7-sonnet-20250219}), The open-source DeepSeek R1 (671B) model.
This selection allows us to assess how {\work} performs across different
model architectures, sizes, and providers.

\subsubsection{Evaluation Dataset: Linux Kernel Driver Analysis with static analyzers}
\label{sec:dataset_rq1} 

Our study utilizes the Android kernel (Linux version 4.14.150, Google Pixel
4XL), which served as the testbed in the original Suture study
\cite{zhang_statically_2021}.
For our initial analysis, potentially informing RQ1 regarding baseline
performance and the challenges of automated bug detection in this complex
environment, we applied prior static analysis tools, Suture and CodeQL-OOB, to the
Linux kernel device drivers.
The key findings from this analysis
provide important context:

\squishlist
    \item \textbf{Suture:} When applied to the Linux device drivers,
    Suture initially generated 251 potential bug reports. This raw
    output translates to a high False Positive (FP) rate of approximately 90\%.
    Suture employed a subsequent semi-automated refinement process, reporting a
    \textit{reviewer-perceived} FP rate of 51.23\%. However, this figure relies on
    Suture's broader definition of a bug, which classified all integer overflows
    as true positives. Furthermore, during our investigation, we identified 4
    additional instances, initially dismissed as FPs by Suture authors' verification,
    that were indeed real bugs (by either their standard or ours).
    This finding highlights potential inconsistencies in large-scale manual
    verification efforts.

    \item \textbf{CodeQL-OOB:}
    We leverage CodeQL \cite{codeql} as a complementary static analysis tool beyond Suture. 
Based on Backhouse et al.'s approach \cite{backhouse_stack_2018}, 
we implemented a simple inter-procedural taint tracking analysis (CodeQL-OOB) 
that traces data flows from \texttt{ioctl} entry points to 
    pointer dereference and \texttt{copy\_from\_user} (two typical cases of 
    OOB bugs) to detect potential stack overflows.
    Running CodeQL-OOB on the Linux device drivers,
    it yields 24 potential bug reports. Our
    manual analysis confirmed 1 of these as a true positive
    (consistent with the known stack overflow bug reported in
    \cite{backhouse_stack_2018}). This corresponds to an FP rate of 95.8\%.
\squishend


\subsubsection{Cost} On average, the cost of running \work is about \$0.1
per case, under the latest version of OpenAI o3-mini.
Each case takes about a few minutes to complete.

\subsection{RQ1: Effectiveness}
\label{sec:eval_rq1}

\begin{table}[h]
    \caption{Performance of \work on top of Suture and CodeQL-OOB.}
    \centering
    \scalebox{.94}{
    \begin{threeparttable}
    \begin{tabular}{lrrrrrrr}
    \toprule
    \textbf{Method} & \textbf{TP} & \textbf{TN} & \textbf{FP} & \textbf{FN} & \textbf{Prec} & \textbf{Rec} & \textbf{F\textsubscript{1}} \\
    \midrule
    Suture  & 24 & 0 & 227 & 0 & 0.10 & - & - \\
    Suture\textsubscript{RP} & 20 & 202 & 25 & 4 & 0.44 & 0.83 & 0.58 \\
    Suture\textsubscript{\work} & 24 & 218 & 9 & 0 & 0.72 & 1.00 & 0.84 \\
    \midrule
    CodeQL-OOB & 1 & 0 & 23 & 0 & 0.04 & - & - \\
    CodeQL-OOB\textsubscript{\work} & 1 & 22 & 2 & 0 & 0.33 & 1.00 & 0.5 \\
    \bottomrule
    \end{tabular}
        
    \end{threeparttable}
    }
    \label{tab:eval}
\end{table}

\subsubsection{Precision and Recall}
Table \ref{tab:eval} shows the performance for the 
evaluated two static analyzers and our post-refinement method, \work.
The results show \work  significantly enhances precision of both Suture (0.10) and  CodeQL-OOB (0.04).
For CodeQL-OOB, \work increases 
precision substantially to 0.33 while not missing any real bugs detected before.
For Suture, the precision is increased to 0.72
due to a drastic reduction in false positives (from 227 to 9).
This refinement does not miss any exisiting bugs. 

Moreover, noting that the semi-automated method in Suture, 
noted as Suture\textsubscript{RP},
actually shows a lower recall (0.83) than the \work refinement (1.0).
This is because after examining the positive results of Suture\textsubscript{\work}, 
we found 4 cases of real bugs that were incorrectly classified as false positives 
during the manual inspection process in Suture\textsubscript{RP}.

\subsubsection{New Bugs} 
As mentioned earlier, we found 4 more cases that are real bugs,
which previously classified as false positive by human inspection in Suture.
Two bugs are from the \texttt{sound} subsystem, reported to the maintainers, 
while waiting for their feedback.
One of them involves a 
data constraint that appears to sanitize a tainted value but can be bypassed 
due to subtle control-flow logic.

The other two bugs are from the \texttt{i2c} subsystem. They involve a condition 
where two tainted values must simultaneously satisfy specific constraints—a case that 
standard taint analyses typically miss due to their focus on single-source propagation.

We have reported all cases following responsible disclosure. Full technical details 
will be made available after the issues are resolved.

\subsubsection{Analysis of FPs}

Despite the general effectiveness of \work, 
it shows 10 FPs.
Upon careful examination of these cases, we attribute 
the inaccuracies to several distinct factors:
\squishlist
\item \textit{Static Analysis Fundamental Limitations} (5 cases): False positives arising from 
inherent limitations in the underlying static analyzers that \work is not designed 
to address. These include imprecisions in taint tracking through complex data structures, 
incorrect indirect call resolution, \etc
\work intentionally operates on the dataflow provided by the static analyzers 
rather than attempting to verify the accuracy of this information itself.
\item \textit{Environment and Language Understanding} (4 cases): Imprecision resulting 
from the LLM's incomplete grasp of C language semantics, hardware-level interactions, 
and kernel-specific programming patterns.
\item \textit{Internal Modeling Errors} (1 case): Inaccuracy originating from a faulty prediction by an internal analysis component (LLM).

\squishend

\subsubsection{Analysis of FNs}
Despite the number of FN shown in the table \ref{tab:eval} is 0, \work still
produced several FNs but gets mitigated by majority voting. Specifically,
we observed that the FNs were concentrated in the \acf{ConA} component, and
the \ac{SecIA} component typically does not generate FNs due to the 
\textit{arbitrary control hypothesis} (AC-Hypo), which will be discussed in
\S\ref{subsubsec:secIA-effectiveness}.

We identified two primary reasons for these FNs:
\squishlist
\item \textbf{Overlooked Complex Conditions:} The LLM sometimes failed to
recognize complex conditions of the program  and 
therefore summarize the pre-/postcondition 
incorrectly.
\item \textbf{Misinterpreting Validation vs. Sanitization:} LLMs occasionally
misclassified sanitization as validations, thereby missing
vulnerabilities.
\squishend

\noindent
These specific failure patterns observed within \ac{ConA} were the direct
motivation for designing the \acf{SAG}.
The \ac{SAG} mechanism enhances the prompts used specifically within the
\ac{ConA} stage, providing targeted instructions aimed at guiding the LLM to
avoid these identified pitfalls (\eg explicitly probing for bypass logic,
carefully differentiating data constraint types).

The positive impact of \ac{SAG} in mitigating these FNs is empirically
demonstrated in RQ2 (\S\ref{sec:eval_rq2}). As shown in
Table~\ref{tab:bug_analysis_results}, the Full Design configuration (using
\ac{ConA} with \ac{SAG} prompts) consistently yields fewer FNs.

Nevertheless, the impact of SAG is not absolute. The ability of a language model
to follow such complex instructions can vary significantly, particularly across
different models. Our results in RQ2/RQ3 (Table~\ref{tab:bug_analysis_results})
indicate that some models could still cause some
residual FNs even with the \ac{SAG}.

\vspace{3pt}
\find{
\textbf{Takeaway 1.} 
\work can effectively post-refine the results of exisiting static analyzers,
hugely improving the precision, and can even find missed bugs.
}

\begin{table}[htbp]
  \centering
  \caption{Bug Analysis Performance Comparison Across LLMs and Design Approaches (Total Cases=120, Real Bugs=22)}
  \label{tab:bug_analysis_results}
  \scalebox{0.89}{
  \begin{tabular}{@{}l| ccc| ccc |ccc@{}}
  \toprule
  \multirow{2}{*}{\textbf{Model}} & \multicolumn{3}{c|}{\textbf{Full Design}} & \multicolumn{3}{c|}{\textbf{w/o \ac{SAG}}} & \multicolumn{3}{c}{\textbf{Simple Prompt}} \\
  \cmidrule(lr){2-4} \cmidrule(lr){5-7} \cmidrule(lr){8-10}
  & FN & FP & F\textsubscript{1} & FN & FP & F\textsubscript{1} & FN & FP & F\textsubscript{1} \\
  \midrule
  OpenAI o3-mini \textcolor{mygold}{\faAward}     & 0  & 3  & \textcolor{mygold}{\textbf{0.94}} & 10 & 1  & 0.67 & 18 & 7  & 0.24 \\
  \midrule
  OpenAI o1 \textcolor{mysilver}{\faAward}       & 3  & 6  & \textcolor{mysilver}{\textbf{0.81}} & 8 & 6 & 0.67 & 18  & 6  & 0.25 \\ 
  OpenAI GPT-4.1 \textcolor{mysilver}{\faAward}   & 1  & 9  & \textcolor{mysilver}{\textbf{0.81}} & 7 & 7 & 0.68 & 3  & 23  & 0.59 \\ 
  Gemini 2.5 Pro      & 12 & 3  & 0.57 & 14 & 4  & 0.47 & 6  & 24 & 0.52 \\
  Claude 3.7 Sonnet   & 13 & 2  & 0.54 & 17 & 2  & 0.34 & 1  & 51 & 0.44 \\
  DeepSeek R1   \textcolor{mybronze}{\faAward}     & 4  & 7  &  \textcolor{mybronze}{\textbf{0.77}} & 10 & 6  & 0.60 & 5  & 42 & 0.42 \\
  \bottomrule
  \end{tabular}
  }
\end{table}

\begin{table}[ht]
    \centering
    \caption{Performance of SecIA, with and without the \acf{AC-Hypo}.}
    \label{tab:secIA-effectiveness}
    \scalebox{0.9}{
    \begin{tabular}{@{}l|cccc|cccc@{}}
    \toprule
    \multirow{2}{*}{\textbf{Model}} & \multicolumn{4}{c|}{\textbf{w/o AC-Hypo }} & \multicolumn{4}{c}{\textbf{w/ AC-Hypo}} \\
    \cmidrule(lr){2-5} \cmidrule(lr){6-9}
    & FP & FN & Prec & Rec & FP & FN & Prec & Rec \\
    \midrule
    OpenAI o3-mini        & 13 & 5  & 0.57 & 0.77 & 38 & 0 & 0.37 & 1.0\\
    \midrule
    OpenAI o1             & 8  & 4  & 0.69 & 0.82 & 39 & 0 & 0.36 & 1.0\\
    OpenAI GPT-4.1             & 15  & 2  & 0.57 & 0.91 & 36 & 1 & 0.69 & 0.95\\
    Gemini 2.5 Pro        & 60 & 3  & 0.24 & 0.86 & 73 & 0 & 0.23 & 1.0\\
    Claude 3.7 Sonnet     & 20 & 2  & 0.50 & 0.91 & 31 & 0 & 0.42 & 1.0\\
    DeepSeek R1           & 25 & 12 & 0.29 & 0.45 & 71 & 4 & 0.18 & 0.82\\
    \bottomrule
    \end{tabular}
    }
\end{table}


\subsection{RQ2: Component Contribution}
\label{sec:eval_rq2}

To address RQ2, we conduct an incremental analysis. This study
evaluates the contribution of the \acf{SecIA}, the
subsequent \acf{ConA}, and the specialized \acf{SAG}
 design used within \ac{ConA}, by comparing
performance across progressively enhanced configurations of \work.

We assess the performance under the following configurations:

\squishlist
\item \textbf{Baseline}: This configuration employs the simple prompt design,
based on related works \cite{wen_automatically_2024,li_iris_2025}.
 It establishes the baseline performance relying primarily on the LLM's inherent capabilities with minimal guidance.
 The baseline design provides a starting point for comparison.

\item \textbf{+ SecIA}: Adds the \ac{SecIA} component to the Baseline. \textit{Purpose:}
 Comparing this to the Baseline isolates the contribution of the \ac{SecIA}
 stage. Detailed metrics for SecIA's filtering rate and soundness are in
 Table~\ref{tab:secIA-effectiveness}.

\item \textbf{+ SecIA + ConA (w/o SAG)}: Adds the \ac{ConA} 
component to the "+ SecIA" configuration, utilizing a simpler prompt design
for constraint checking (\ie without SAG). Comparing this to ``+
 SecIA'' isolates the contribution of adding the constraint assesses step
 itself.

\item \textbf{Full Design (+ SecIA + ConA + SAG)}: This configuration enhances
 the \ac{ConA} component from the previous step by incorporating the specialized
 \ac{SAG} design. This represents the complete
 \work{} system. Comparing this to previous configurations 
 emphasize the contribution of the \ac{SAG}.
\squishend

\noindent
For this component analysis (RQ2) and the subsequent model versatility analysis
(RQ3), we focus our evaluation on a dataset derived from Linux kernel analysis,
specifically targeting the \texttt{sound} module. This module was selected
because the original Suture study identified it as containing a high density of
true positive vulnerabilities (22 out of 24 known bugs), providing a rich testbed for
assessing bug detection capabilities.
The dataset consists of 120 cases, with 22 known bugs (positives) and 98
non-bug cases (negatives).

We evaluate the performance of these components for diverse LLMs, including
OpenAI's o3-mini, o1,  GPT-4.1, Gemini 2.5 Pro, Claude 3.7 Sonnet, and DeepSeek R1. The
overall performance results for these configurations are summarized in
Table~\ref{tab:bug_analysis_results}, while Table~\ref{tab:secIA-effectiveness}
provides the detailed breakdown specifically for the \ac{SecIA} component's
effectiveness and filtering metrics.

\subsubsection{Baseline}

Our experimental results clearly demonstrate the significant contribution of our
proposed multi-phase workflow and its components compared to a baseline
approach.
As shown in Table~\ref{tab:bug_analysis_results}, despite some models like
Claude 3.7 Sonnet (FN=1) and GPT-4.1 (FN=3) showed low False Negatives,
potentially reflecting their raw analytical power, this came at the cost of high
False Positives (FP=51 and FP=23, respectively), rendering this simple design
ineffective for practical use. The F1 scores for the baseline were generally low
across models. This direct prompting approach demonstrated worse performance
when compared to other \work configurations.

\subsubsection{\acf{SecIA}}
\label{subsubsec:secIA-effectiveness}

As Table \ref{tab:secIA-effectiveness} shows, 
our Arbitrary Control Hypothesis (AC-Hypo)
enhances recall across all models.
 Without AC-Hypo, the models
exhibit noticeable False Negatives, ranging from 2 to 12 FN cases across
tested models. After applying AC-Hypo, the FN rate decreases to
zero for all models except DeepSeek R1 and GPT-4.1, which maintains a low FN rate (4 and 1, respectively),
achieving a high recall of 0.82 and 0.95.

Meanwhile, SecIA demonstrates strong effectiveness as a security vulnerability \textit{filter}. 
Taking OpenAI's
o3-mini as an example, among a total of 98 negative cases, SecIA
successfully filters out 60 cases (TN). 
This indicates that SecIA not only has a high recall rate,
but it is also effective, substantially improving analysis efficiency.

\subsubsection{\acf{ConA} without \acf{SAG}} 
While this multi-phase workflow (\ie SecIA + ConA) significantly
reduces the high volume of FPs seen in the
Baseline; for instance, Claude 3.7 Sonnet's FPs dropped from 51 to 2, and
Gemini 1.5 Pro's from 24 to 4. It also leads to a significant increase
in False Negatives (FNs) for Gemini 2.5 Pro (FN=14),  Claude 3.7 Sonnet
and DeepSeek R1 (FN=17), and OpenAI o1 (FN=10) in the `w/o SAG'.
This supports our hypothesis
(\S\ref{subsubsec:naive-llm}) that providing constraints, while helpful for
pruning obvious non-bugs, can encourage LLMs to become overly confident. Once
patterns suggesting data validity are identified, the LLM may default to
classifying the issue as ``not a bug,'' reflecting a potential statistical bias
towards common safe patterns rather than performing nuanced reasoning about
subtle flaws or bypass conditions.

\subsubsection{\acf{SAG}}
Comparing the Full Design (using SAG within ConA) to the w/o SAG configuration
in Table~\ref{tab:bug_analysis_results} demonstrates SAG's effectiveness.
Introducing SAG leads to a substantial reduction in FNs across all tested
models. Consequently,
the overall F\textsubscript{1} score sees a marked improvement with SAG (e.g.,
improving from 0.67 to 0.94 for o3-mini and 0.34 to 0.54 for Claude). This
indicates that \ac{SAG} successfully guides the LLM within \ac{ConA} to overcome
the previously observed overconfidence, achieving a better balance between FP
reduction and FN mitigation.

\vspace{3pt}
\find{
\textbf{Takeaway 2.} 
The design components of \work enables effective LLM bug analysis by
significantly reducing both FP and FN compared to baseline prompting.
}



\subsection{RQ3: Model Versatility}
\label{sec:eval_rq3}

The results shown in Table \ref{tab:bug_analysis_results} 
affirm that \work is a general LLM-based technique applicable across different models,
consistently improving upon baseline performance. However, the degree of success
highlights variations in how different LLMs interact with complex instructions
and structured reasoning processes.

As noted in RQ2, the baseline performance  offers a glimpse
into the models' raw capabilities, somewhat correlating with general LLM
benchmarks where Gemini 2.5 Pro and Claude 3.7 Sonnet are often considered
leaders \cite{vellum2025leaderboard}. However, this raw
capability did not directly translate to superior performance within our
structured task without significant guidance.

When employing the `Full Design' of \work, we observed distinct differences in
instruction-following adherence. The OpenAI models, o1, GPT-4.1 (F1=0.81), and
particularly our core model o3-mini (F1=0.94), demonstrated excellent alignment
with the workflow's intent. 

Conversely, while the `Full Design' significantly improved the F1 scores for
Gemini 2.5 Pro (0.57) and Claude 3.7 Sonnet (0.54) compared to their baseline
or `w/o sag' results by drastically cutting down FPs, they still struggled with
relatively high false negatives (FN=12 and FN=13, respectively). This suggests
that even with the \ac{SAG}, these powerful models may face
challenges in precisely balancing the various analytical steps or interpreting
the nuanced instructions within our workflow, possibly still exhibiting a degree
of the over-confidence (for ``sanity check'')
that \ac{SAG} could not fully overcome in their case. DeepSeek R1 (F1=0.77)
showed a strong, balanced improvement, landing between the GPT models and the
Gemini/Claude in terms of final performance with the full design.
This demonstrates the generality of the \work to open-sourced models.

In summary, while our approach is broadly applicable, its optimal performance
depends on the LLM's ability to robustly follow complex, multi-step
instructions, with models like OpenAI's o3-mini currently showing the strongest
capability in this specific structured bug analysis task.

\vspace{3pt}
\find{
\textbf{Takeaway 3.} \work shows
broad applicability and yields promising results across diverse LLMs, including
the open-source DeepSeek R1. OpenAI's o3-mini currently gets the best
result.
}



\subsection{Case Study: Data Structure Traversal}
\label{apdx:traversals}

Linux kernel code often uses pointers to traverse data
structures. For example, the
following code walks through a linked list \texttt{list} using a marco
\texttt{list\_for\_each\_entry}:

\begin{center}
\includegraphics[width=\linewidth]{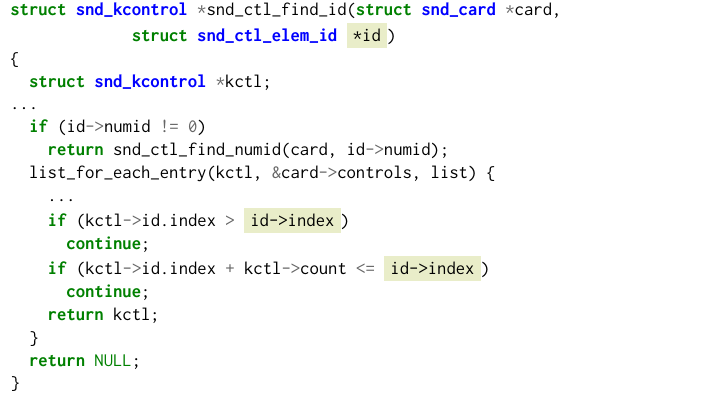}
\end{center}




In this code, the loop goes through each element in the list and checks if any
of them match the input id, which comes from the user.
This kind of loop is hard for static analysis tools to understand unless
they are specially designed to handle linked lists. These tools might wrongly
report a warning, saying that user input affects when the loop ends. But in
reality, the loop always stops at the end of the list, not because of the user
input.

The user input only decides which element gets picked, not how long the loop
runs. Our method, \work, avoids this false warning by using the LLM's deeper
understanding of how data structures like linked lists work.

\subsubsection{A Bypassable Condition}

There's also a tricky part in this code. Before 
the \texttt{list\_for\_each\_entry} loop even runs, 
there is a condition that checks
if the \texttt{id->numid} is not zero. If it is not zero, the function will return
the result of \texttt{snd\_ctl\_find\_numid(card, id->numid)}. 

This means  the checks inside the loop for \texttt{id->index} might not happen at all.
So even the loop contains conditions seem to validate \texttt{id->index},
those checks is not always are only performed if \texttt{id->numid} is zero.
Otherwise, the function skip the loop entierly.

This cases directly causes false negatives in \work, especially without the 
\ac{SAG} mechanism.
Limited by space, we describe this case in detail in the appendix \S\ref{apdx:traversals-2}.



%% file: sec/end_nominted.tex
\section{Limitation \& Discussion}
\label{sec:limitation}


\textbf{Dependency on LLMs.}
The performance of \work is influenced by the capabilities of the underlying LLMs used for reasoning.
While this dependency could potentially impact external validity, our experiments demonstrate \work's 
robustness across different models.
We have evaluated \work with multiple LLMs, including both closed-source and open-source models like Deepseek R1, 
which achieves approximately 80\% of the performance of top-performing models.
These results confirm that \work's approach generalizes well across different models, 
though performance variations exist.

\noindent
\textbf{Limited Checkers and Bug Scope.}
Our evaluation uses only taint-style checkers—those inherited from 
and Suture (and Dr.checker), 
plus one CodeQL port—focused on memory-safety and DoS
bugs in the Linux kernel. This narrow setup threatens external validity,
and the performance we report may not carry over to
other bugs (\eg data races) or analysis frameworks. Although the extra
CodeQL checker suggests \work\ can transfer across tooling, a broader study is
needed for general applicability.

\noindent
\textbf{Towards More Sound Analysis.} The soundness of current implementation
 could be improved through two directions: (1) Adding symbolic verification
 \cite{bhatia_verified_2024, cai_automated_2025} to validate the LLM's reasoning and refine outputs
 based on formal methods (2) Implementing a hybrid architecture where the LLM
 performs initial code slicing while symbolic execution handles constraint
 analysis, combining the LLM's contextual understanding with provably sound
 formal reasoning.

\noindent \textbf{Intergation with More Static Analysis Tools.}
The CodeQL-OOB is an interesting static analysis tool that 
is based on the lightweight static analysis framework, CodeQL.
Considering that \work can effectively improve the precision of existing static analysis tools,
we could implement more corsare-grained (with low precision) static analysis tools
and integrate them into \work.
Compared to heavy static analysis tools such as suture, an imprecise static analysis
could be much easier to implement and maintain.


\section{Related work}
\label{sec:related}

\textbf{LLM for Program Analyses \& Bug Detection}
LLMs have been widely applied to program analysis tasks for bug detection.
IRIS \cite{li_iris_2025} combines LLMs' contextual understanding with CodeQL
queries to enhance taint analysis. LLM4SA \cite{wen_automatically_2024} 
leverages LLMs to refine static analysis results. Their approach serves as a 
baseline for \work (referred as simple prompt).
LLMDFA and LLMSAN~\cite{DBLP:conf/emnlp/WangZSX024,DBLP:conf/nips/WangZSXX024}
use LLMs to perform data flow analysis and bug detection. 
LLift \cite{llift} focuses on use-before-intialization bugs and prompts 
LLMs to identify and summarize possible initializers.
Focusing on real-world problems,
\work upgrades the scope to a general taint-style bug detection cross multiple
functions.

\noindent
\textbf{LLMs for program analysis.}
LLMs have been widely applied to program analysis tasks, including static
semantics analysis~\cite{DBLP:conf/nips/WangZSXX024,
DBLP:journals/corr/abs-2501-18160}, indirect call
resolution~\cite{cheng_semantic-enhanced_2024}, and various 
 inference tasks in program verification and synthesis 
 ~\cite{chow_pyty_2024, le-cong_can_2025, pei_can_2023, misu_towards_2024,
cai_automated_2025,li_iris_2025}. Similarly, \work leverages LLMs to reason about
security impacts, utilizing their contextual understanding of code semantics to
enhance the precision of program analysis results.


\noindent
\textbf{Reasoning for LLMs.}
Despite their success on many tasks, the ability of LLMs to reason about code semantics and behaviors remains an active area of research~\cite{ding2024semcoder, zelikman2022star, li2025codei, ni2024next}. 
Recent studies have shown that LLMs are still far from performing reliable code reasoning, and their predictions are thus fragile and susceptible to superficial changes in input~\cite{pei2023exploiting, steenhoek_err_2025, hochlehnert2025sober}. 
This fragility is often attributed to the learned models taking ``shortcuts'' based on superficial patterns in training data rather than robust, generalizable reasoning strategies~\cite{mccoy_embers_2024, yefet2020adversarial, gao2023discrete, yang2022natural, bielik2020adversarial}.
\work mitigates this problem with a similar spirit to existing works on boosting the LLMs' reasoning by constraining their reasoning space with structural and symbolic procedures~\cite{li2025structured, chen2022program, li2023chain, chen2024steering, yao2023tree}.


\section{Conclusion}
\label{sec:conclusion}
This paper introduces \work{}, an innovative post-refinement framework that
integrates Large Language Models (LLMs) with static analysis. By employing
\acf{SecIA}, \acf{ConA}, and \acf{SAG} to guide LLMs through the reasoning process,
\work{} significantly enhances the
precision of initial static analysis findings without sacrificing scalability.
Our evaluation demonstrates that \work{} dramatically reduces false positives in
Linux kernel Analysis, minimizes
manual inspection effort, and uncovers previously ignored vulnerabilities,
highlighting the promise of guided LLMs in making automated bug detection more
practical and effective.

%% file: sec/appendix_nominted.tex
\appendix
\section{Case Study of \work}
\label{apdx:case-study}

\subsection{Data Structure Traversals \& Bypass Conditions}
\label{apdx:traversals-2}


In \S\ref{apdx:traversals}, we have shown a case of data structure traversal:

\begin{center}
\includegraphics[width=\linewidth]{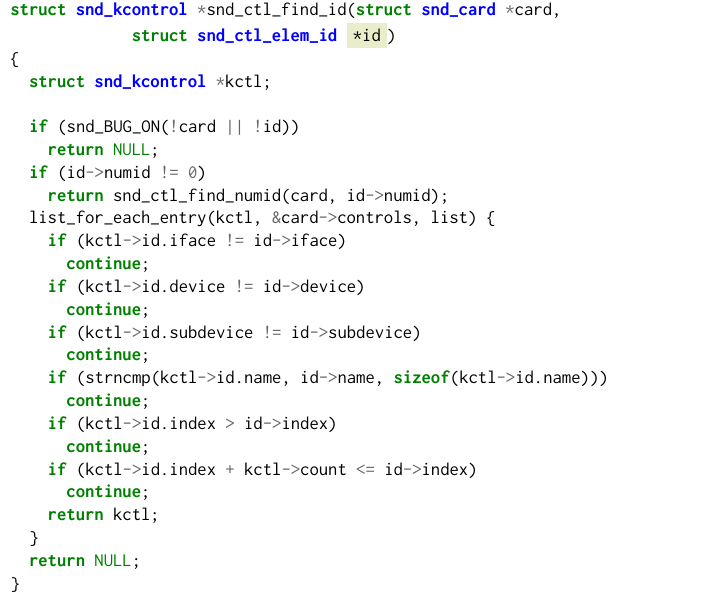}
\end{center}





We now consider a caller of our example \texttt{snd\_ctl\_find\_id}.
The function \texttt{snd\_ctl\_elem\_write} takes a
\texttt{control} parameter, which is also tainted user input. It then calls
\texttt{snd\_ctl\_find\_id} to find the corresponding \texttt{kctl} object.
The \texttt{put} function of \texttt{kctl} is then called with the tainted
\texttt{control} parameter.

\begin{center}
\includegraphics[width=\linewidth]{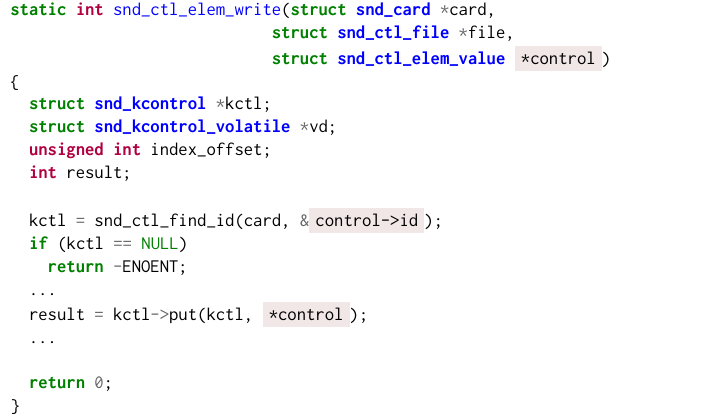}
\end{center}

\subsubsection{Step 2: Constraint Collection}

Suppose the \texttt{put} function is a sink or critical operations that could
lead to a vulnerability (and there are indeed many such cases, for different 
instances of \texttt{kctl}). 

When \acf{ConA} analyzes this code, its second step is to collect the 
data constraints of for the tainted input \texttt{control}
(actually, the tainted input is \texttt{control->id}, and specific field depends on how the exactly \texttt{put}
uses). 

Natrually, it will collect the following constraints: 
\texttt{snd\_ctl\_find\_id(...)} cannot return \texttt{NULL}.
and the \texttt{kctl} object must be valid.

\subsubsection{Step 3: Constraint Effect Summarization}
After collecting the constraints, the next step is to summarize the effect of
the constraints, so we need to go back to the \texttt{snd\_ctl\_find\_id} function
and analyze the constraints of \texttt{control->id}.

Here's the most triky part, suppose we track the taint data \texttt{control->id.index}
and by looking at the definition of \texttt{snd\_ctl\_find\_id}, we can see that:

\begin{center}
\includegraphics[width=\linewidth]{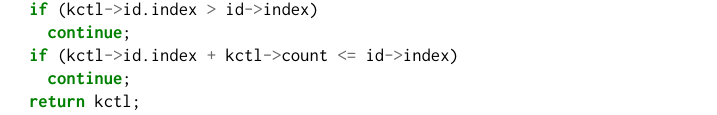}
\end{center}

Naively, LLM may think these two conditions ensures the \texttt{control->id.index} is
within a strict range, between \texttt{kctl->id.index} and \texttt{kctl->id.index + kctl->count\texttt}
Considering that these \texttt{kctl} objects are predefined and maintained
 in the kernel, we might think our tainted input is used with an 
 effective and restricted range.

\noindent
\textbf{A Closer Look:} However, this is not the case. The key here is before
the \texttt{list\_for\_each\_entry} loop:

\begin{center}
\includegraphics[width=\linewidth]{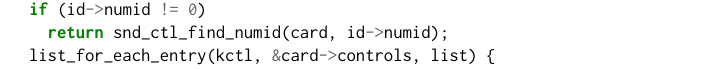}
\end{center}

Before the loop, the function checks if \texttt{id->numid} is not zero, and if so,
it calls \texttt{snd\_ctl\_find\_numid} to find the corresponding \texttt{kctl}
object, and returns it directly. This means that the \texttt{list\_for\_each\_entry}
loop, and inside checks can be \textit{\textbf{bypassed}} if the \texttt{id->numid} is not zero.

The function \texttt{snd\_ctl\_find\_numid} is defined as:
\begin{center}
\includegraphics[width=\linewidth]{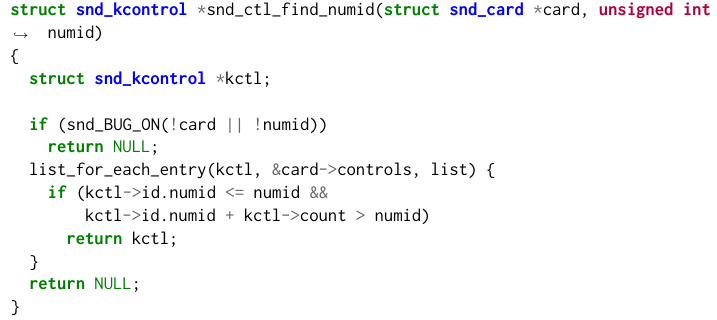}
\end{center}

This function also traverses the \texttt{card->controls} list, but it uses
\texttt{kctl->id.numid} as the key to find the corresponding \texttt{kctl}
object. The \texttt{numid} is a unique identifier for each \texttt{kctl} object,
and it is not directly related to the \texttt{control->id.index} field.

Finally, the current effect summarization for the \texttt{snd\_ctl\_find\_id} function
should be like (not shown completely):

\begin{enumerate}
    \item Precondition: \texttt{control->id.numid != 0},
    
    Postcondition: \texttt{kctl != NULL \&\& kctl->id.numid <= control->id.numid < kctl->id.numid+kctl->count}
    \item Precondition: \texttt{control->id.numid == 0}
    
    Postcondition: \texttt{kctl != NULL \&\& kctl->id.index <= control->id.index < kctl->id.index+kctl->count}

\end{enumerate}

The overlook of this bypass is the main reason of the false negatives 
for models.

\section{Outline of LLM Prompt}
\label{apdx:prompt}

We provide the outline of the LLM prompt for each task in \work.
For simplified, we remove all examples (that used in few-shot in-context learning).

The \texttt{args} are the arguments for each case running,
and the \texttt{callback} is the callback function for each prompt for the agentic
    design PKA.

\subsection{Prompt for Inferring Variable Names}

\begin{lstlisting}[numbers=none]
I have a static analysis tool that tracks tainted user input in Linux kernel drivers.
Since the analysis uses LLVM IR, I need help identifying the corresponding source-level variables.
For each case:

1. Tainted values are either local variables loaded from globals/parameters, or direct function parameters
2. Tainted values are always numeric
3. Be aware of name redundancy, especially in global variables and their fields
4. Identify specific struct field names when applicable
5. For propagation, identify the first local variable receiving the taint

Analyze line by line:
  ========================================
  The bug detector is: {}
  source code:
  {}
  ---------------------------------------
  Instructions:
  {}
  ---------------------------------------
  line no (in source code): {}
  ---------------------------------------
args:
  - get_bug_detector
  - get_function_first_part
  - get_insts_from_ctx
  - get_source_line_set
\end{lstlisting}

\subsection{Prompt for SecIA}

\begin{lstlisting}[numbers=none]
I have a static analysis tool for Linux kernel drivers that produces many false positives when detecting:
1. Tainted Arithmetic Operations (integer overflows)
2. Tainted Loop Bound Conditions (infinite loops, unexpected iterations)
3. Tainted Pointer Dereferences (arbitrary memory access)
4. Buffer Overflow (out-of-bounds access)
5. Tainted length in copy_from_user (especially stack overflow)
Your Task:
1. Analyze the provided code snippet flagged by our tool
2. Assume:
    - Attackers cannot control the kernel
    - Tainted variables can be set to any value within their type range
    - All existing checks can be bypassed
    - Ignore any security checks in the code
3. Determine if the code represents:
    - Potential Bug: If the tainted variable can cause infinite loops, very large loops, or memory bugs in the current context
    - Normal Code: If the tainted variable usage doesn't lead to security issues within our scope
Provide a step-by-step explanation of your reasoning and classification.
========================================
tainted_varaible: {}
bug detector: {}
source code 
{}
========================================

args:
  - get_tainted_value
  - get_bug_detector
  - get_function
\end{lstlisting}

\subsubsection{Schema-constrained summarization}

\begin{lstlisting}[numbers=none]
Summarize our discussion, and respond with a <bug_eval> tag indicating whether the tainted variable can lead to vulnerabilities:
currently we only consider (1) infinite/very-large loop, (2) out-of-bound access/buffer overflow/arbitrary memory access;
for other types of bugs, you can say "normal code" or "not_a_bug"

<bug_eval>
    <tainted_var>tainted_var</tainted_var>
    <vulns>
        <vuln>
            <type>out_of_bound_access</type>
            <desc>Brief description of how the vulnerability occurs</desc>
        </vuln>
    </vulns>
</bug_eval>

For no vulnerability: <bug_eval>not_a_bug</bug_eval>
For potential vulnerability: <bug_eval>potential_bug</bug_eval>
If uncertain: <bug_eval>uncertain</bug_eval>
\end{lstlisting}

\subsection{Prompt for ConA}

This part we provide the prompt for the \acf{ConA} task.
These prompts already contain the design of \ac{SAG}.
Additionally, we also provide a list of functionalities of the 
agent \ac{PKA} inside prompts, represented as 
\colorbox{sageleaf}{\texttt{\{AGENT PROMPTS HERE\}}}.

\subsubsection{Step 1: Reachability Analysis}

The following is the prompt for the step 1 of \ac{ConA} task.
(see \S\ref{subsubsec:san-step1} for details)

\vspace{0.5em} 
\begin{lstlisting}[numbers=none]
Identify the sink (the last line of the provided function context) and determine its preconditions.
Focus only on two types of preconditions:
1. Direct checks (dominate conditions) - Conditions that directly control if the sink executes
   Example: `if (flag) { sink(tainted_var); }`
2. Early returns/bypasses (guard conditions) - Conditions that cause early returns or jumps that bypass the sink
   Example: `if (tainted_var > 100) return; sink(tainted_var);`
   Here, "tainted_var <= 100" is the precondition to reach the sink.
Ignore conditions that don't directly impact sink reachability.
========================================
<@\colorbox{sageleaf}{\{AGENT PROMPTS HERE}\}@>
========================================
sink variable
{}
sink context (the full context of the function):
{}
========================================
Summarize in this format:
```xml
<sink_precondi>
 <precondi>
  <type> dominate_condition </type>
  <condition> flag </condition>
  <dominated_sink> if(flag) sink(tainted_var) </dominated_sink>
 </precondi>
 <precondi>
  <type> guard_condition </type>
  <condition> tainted_var <= 100 </condition>
  <guard_bypass> if (tainted_var > 100) return/goto invalid_label; </guard_bypass>
 </precondi>
</sink_precondi>
```
Note: Multiple preconditions combine with "AND" logic.
  
args:
    - get_tainted_value
    - get_function
callback:
    - need_struct_def
    - need_global_var_def
\end{lstlisting}

\subsubsection{Step 2: Constraint Collection}

The following is the step 2 of the prompt for the \ac{ConA} task.
(see \S\ref{subsubsec:san-step2} for details)

\begin{lstlisting}[numbers=none]
Help find range constraints for the tainted variable across the call chain.
========================================
tainted_variable: {}
current callchain: {}
context of the sink:
{}
========================================
Focus on finding all possible value range constraints for the tainted variable, without analyzing their effectiveness.
Types of constraints to look for:
1. Validation: conditions that reject invalid ranges (e.g., if (tainted_var < 0) return -EINVAL;)
2. Sanitization: corrections applied to the value (e.g., tainted_var = min(tainted_var, 100);)
3. Type constraints: implicit limits from variable types (e.g., uint8_t limits to [0,255])
Important notes:
- Validations are transferable through operations (e.g., constraints on var = tainted_var + 1 apply to tainted_var)
- Sanitizations are not transferable
- Base your analysis only on the provided code, not prior knowledge
- Track the variable across different names in different contexts
=========================================
<@\colorbox{sageleaf}{\{AGENT PROMPTS HERE\}}@>
=========================================
Summarize your findings in this format:
---------------------------------------
<range_constraints>
  <constraint>
    <type>validation|sanitization|type_constraint</type>
    <handler_func>function_name</handler_func>
    <context>relevant_code_snippet</context>
  </constraint>
</range_constraints>
---------------------------------------
args:
  - get_tainted_value
  - get_call_chain
  - get_function_first_part
callback:
  - need_func_def
  - need_caller
  - need_struct_def
  - need_global_var_def
\end{lstlisting}

\subsubsection{Step 3: Constraint Effect Summarization}

The following is the prompt of the step 3 of the \ac{ConA} task.
(see \S\ref{subsubsec:san-step3} for details)

\begin{lstlisting}[numbers=none]
Act as a program verifier/symbolic execution engine to infer precondition and postcondition pairs for constraints in code. Use these simple rules:
- The **precondition** is the path condition required to reach the constraint.
- The **postcondition** is the effect on the tainted variable (often shown as its valid range).

Consider all branches and early return "bypass" cases. For instance, given this sample function:
```c
void check_tainted_value(int config, int other_config, int x){
  if (other_config == CHECK_SKIP)
    return;
  if (config == 1)
    assert(x > 0 && x < 100);
  else if (config == 2)
    assert(x > 0);
}
You must extract the following [precondition, postcondition] pairs:
1. Bypass Case
  Precondition: other_config == CHECK_SKIP
  Postcondition: x in (-inf, +inf)
2. Config 1 Case
  Precondition: other_config != CHECK_SKIP && config == 1
  Postcondition: x in (0, 100)
3. Config 2 Case
  Precondition: other_config != CHECK_SKIP && config == 2
  Postcondition: x in (0, +inf)
4. Default Case
  Precondition: other_config != CHECK_SKIP && config != 1 && config != 2
  Postcondition: x in (-inf, +inf)

========================================
<@\colorbox{sageleaf}{\{AGENT PROMPTS HERE\}}@>
=======================================

  Finally, output your analysis as XML in the following format:
```xml
 <range_constraint>
   <type>validation</type>
   <handler_func>check_tainted_value</handler_func>
   <condition_pairs>
   <pair>
     <precondi>other_config == CHECK_SKIP</precondi>
     <postcondi>x in (-inf, +inf)</postcondi>
     <context> if (other_config == CHECK_SKIP) return; </context>
   </pair>
   ...
   <pair>
     <precondi>other_config != CHECK_SKIP && config != 1 && config != 2</precondi>
     <postcondi>x in (-inf, +inf)</postcondi>
     <context> induced by the other branches </context>
   </pair>
   </condition_pairs>
 </range_constraint>
````
callback:
  - need_func_def
  - need_struct_def
  - need_global_var_def
  - need_caller
\end{lstlisting}

\subsubsection{Step 4: Final Vulnerability Evaluation}

The following is the prompt of the step 4 of the \ac{ConA} task.
(see \S\ref{subsubsec:san-step4} for details)

\begin{lstlisting}[numbers=none]
Evaluate if this bug is eliminated, not exploitable, or still vulnerable.
Extract concrete values for "size of" or "length of" in constraints and bug conditions.
Analyze whether the bug condition can be satisfied considering all constraints:
  - Disregard constraints with postconditions that don't limit the tainted variable
  - Disregard constraints for unrelated sinks
  - For kernel-controlled conditions, determine reachability based on your knowledge
  - If a precondition contains user-controlled variables, assume users can bypass it
A bug is "eliminated" only if:
  - The postcondition restricts the tainted variable to a safe range that makes the bug impossible
  - The precondition is always satisfied when the sink's precondition is true
If the postcondition isn't strong enough, assess exploitability assuming attackers can set any value.
========================================
<@\colorbox{sageleaf}{\{AGENT PROMPTS HERE\}}@>
========================================

callback:
  - need_func_def
  - need_struct_def
  - need_global_var_def
  - need_caller
\end{lstlisting}

\subsubsection{Schema-constrained summarization}
The following is the prompt for summarization of the \ac{ConA} task,
which is used to extract the final result of the vulnerability evaluation
within a \texttt{<final\_res>}.

\begin{lstlisting}[numbers=none]
Now Let's summarize our discussion, and respond in a <final_res> tag with the following format:
"still_a_bug", "eliminated", "likely_safe", "likely_unsafe", "not_exploitable" or "uncertain" within a <final_res> tag, e.g., <final_res>still_a_bug</final_res>
\end{lstlisting}

\subsection{Prompt for PKA}

The following is the exact prompt that shown
as \texttt{AGENT PROMPTS HERE} in the previous parts.

\begin{lstlisting}{numbers=none}
========================================
First of all, you don't need to complete the task in your initial response. You can always ask for more information.

When you need additional details, use the following format. In this case, don't reach a conclusion immediately - instead, request the information you need to perform a thorough analysis once you receive my response.
--- request 1: ask for the function definition ---
You could ask me for the definition of the function. in this case, you could respond with the following:
<requests>
  <request>
    <name>need_func_def</name>
    <args>
      <arg>func_1</arg>
      <arg>func_2</arg>
    </args>
  </request>
</requests>
--- request 2: ask for the struct definition ---
You could ask me for the definition of the structure. in this case, you could respond with the following:
<requests>
  <request>
    <name>need_struct_def</name>
    <args>
      <arg>struct_name_1</arg>
      <arg>struct_name_2</arg>
    </args>
  </request>
</requests>
--- request 3: ask for the caller of the current function ---
You could ask me for the caller for the current function. (Note: you can only request one caller at a time)
 in this case, you could respond with the following:
 <requests>
  <request>
    <name>need_caller</name>
    <args>
      <arg>current_function_name</arg>
    </args>
  </request>
</requests>
--- request 4: ask for the definition of global variables ---
You could ask me for the definition of global variables. In this case, you could respond with the following:
<requests>
  <request>
    <name>need_global_var_def</name>
    <args>
      <arg>global_var_1</arg>
      <arg>global_var_2</arg>
    </args>
  </request>
</requests>
========================================
\end{lstlisting}